\documentclass[referee]{aa}
\usepackage{graphicx}
\begin{document}
%\thesaurus{}
\title{Large scale star formation in galaxies. \break
II. The spirals NGC 3377A, NGC 3507 and NGC 4394.}
\subtitle{Young Star Groupings in Spirals}
\author{A. Vicari$^1$, P. Battinelli$^2$, R. Capuzzo--Dolcetta$^1$, T.K. Wyder$^3$ and G. Arrabito$^4$}

\institute {
$^1$Dipart. di Fisica, Universit\`a La Sapienza, P.le A. Moro 5, Roma - Italy\\
$^2$Oss. Astronomico di Roma, Viale del Parco Mellini 84, Roma - Italy\\
$^3$California Inst. of Technology MC 405-47 1200 E. Cal. Blvd Pasadena, CA 91125\\
$^4$Dipart. di Matematica, Universit\`a La Sapienza, P.le A. Moro 5, Roma - Italy\\
}
\offprints{alessandrovicari@uniroma1.it}

\date{Received; accepted}

\authorrunning{Vicari et al.}
\titlerunning{Young Star Groupings in Spirals}

\abstract{
The identification of young star groupings (YSG) in the three spiral galaxies
NGC 3377A, NGC 3507, NGC 4394 is obtained by mean of the statistical
method described in Paper I. We find 83, 90, 185
YSGs, respectively. An identification map of YSGs,
as well as their size distribution, their
B-luminosity function, their surface luminosity density radial
behaviour, are presented and comparatively discussed.
These data, in addition to those in Paper I, constitute a first sample suitable
for seeking correlations among properties of galaxies and their YSGs, which we
briefly discuss here.
}

\maketitle

\section{Introduction}
Global star formation in a spiral galaxy is influenced by several factors.
Different authors have investigated the dependence of the star
formation rate and efficiency on environment.
Star formation in interacting galaxies is generally enhanced,
while it is not well assessed its dependence
upon galaxy morphological type (Young et. al
1996 and references therein).
Some recent papers have shown the importance of the determination of empirical
correlations between large scale star forming regions and parent galaxy
properties in understanding the star formation process.
In this framework Elmegreen et al. (1994, 1996) and
Elmegreen \& Salzer (1999), analyzing different samples of spirals, have found
a clear correlation between the size, the total B magnitude of a galaxy
and the size of its largest stellar complex.
These authors suggest, indeed, that these scaling laws can be related to the
properties of the interstellar medium and of the instability process itself.
However, it is worth to note that Selman \& Melnick (2000) have recently explained that
the correlation between the size of the largest superassociation and the
luminosity of the parent galaxy, found by Elmegreen et al. (1994), is most likely
due to a size-of-sample effect.

The problem of the identifications of regions of star
formation in distant, unresolved, galaxies is quite difficult, because the
identification suffers from several biases, like
those introduced by differences in observational data and/or
in the identification criterion adopted.
For instance, Hodge (1986) clearly discussed these difficulties when trying
to derive general properties of star forming regions.

To overcome these complications we developed an automatic method for the
identification of the star forming regions (Young Star Groupings, or YSGs
hereafter) in unresolved galaxies (Adanti et al. 1994). The availability of
various sets of colours and fluxes allows every pixel of the galaxy image to
be represented as a point in the space of the variables (i.e. of fluxes and colours).
Principal Component Analysis (PCA) and Cluster Analysis (CA) result in an artificial
image of the galaxy, where pixels are grouped into classes according to their relative
distances in the space of the variables. As in Battinelli et al. (2000,
hereafter Paper I), we performed our
classification using U, U-B, B-V and B-R as variables.
A first step
to a deeper understanding of the link between the YSG's and parent galaxy's
properties is to build a homogeneous and \lq objective\rq~ database.
Target galaxies were chosen according to  the criteria discussed in Paper I.

In this paper we present data and results of YSGs in
three spiral galaxies: NGC 3337A, NGC 3507 and NGC 4394.
In Sect. 2 observational data acquisition and reduction are described;
in Sect. 3 we present and discuss our identifications in each galaxy.
Finally, in Sect. 4, we discuss some properties of the sample of YSGs in all
the six galaxies we have studied so far.

\section{Observations and data reduction}
All data were obtained with the SPIcam camera at the 3.5 m telescope
of the Apache Point Observatory (New Mexico, USA)
\footnote{APO is privately owned and operated by the Astrophysical
Research Consortium (ARC), consisting of the University of Chicago, the
Institute for Advanced Study, Johns Hopkins University,
New Mexico State University, Princeton University, the University of
Washington and Washington State University.},
in four different broadbands: U, B, V and R. SPIcam is a backside
illuminated thinned SITe CCD,
with a size of $1024\times1024$ pixels, with an angular scale of
$0.28\arcsec$/pixel and with a field of view of about $4.8 \arcmin \times4.8\arcmin$.

Three exposures per filter were taken for each galaxy. The CCDPROC
routine in IRAF\footnote{IRAF is a free software distributed by
National Optical Astronomy Observatories, which are operated by the
Association of Universities for Research in Astronomy Inc., under
contract to the National Science Foundation.} was used to subtract
the bias, as measured in the overscan region of each image, as well
as to divide by the appropriate flat field for that night and filter.
The spatial offsets among the images of a particular galaxy were obtained using
the positions of stars within each field. The images were shifted to
a common reference frame using the IMALIGN routine in IRAF. Finally,
the IRAF task IMCOMBINE was used to average together the three exposures
in each filter while at the same time rejecting cosmic rays.

NGC 3377A was observed on the night of 1998 November 21, a photometric
night, while the NGC 3507 and NGC 4394 images were taken on 1998
December 21 and on 17-18 January 1999 respectively. These last three
nights were not photometric, so we obtained shallower exposures of NGC
3507 and of NGC 4394 on a clear night, 16 April 1999, which we used to
calibrate these images. The ratio in each filter between the averaged
images of the non-photometric nights and the shallower images taken on
the last photometric night was determined using stars within the
field, if present, or using the radial profile of the galaxy itself.
Total exposure times are always 1200 s in the U images and 600 s in
the B, V and R bands, with the exception of the U image of NGC 4394,
whose exposure time is 2000 s. The seeing of each image is reported in
Table 1.

\begin{table}[ht]
\begin{center}
\caption{Observational log}
\begin{tabular}{|l|c|c|c|c|c|}\hline\hline

			&\textsf{Filter}	&\textsf{Date}		&\textsf{Exp. time}	&\textsf{FWHM}\\
&			&\textsf{(d/m/y)}	&\textsf{(sec)}		&\textsf{(arcsec)}\\\hline

			\textsf{NGC 3377A}	&\textsf{U}		&\textsf{21/11/98}	&\textsf{3$\times$400}	&\textsf{1.18}\\
			&\textsf{B}		&\textsf{21/11/98}	&\textsf{3$\times$200}	&\textsf{1.18}\\
			&\textsf{V}		&\textsf{21/11/98}	&\textsf{3$\times$200}	&\textsf{1.40}\\
			&\textsf{R}		&\textsf{21/11/98}	&\textsf{3$\times$200}	&\textsf{1.18}\\\hline

			\textsf{NGC 3507}	&\textsf{U}		&\textsf{21/12/98}	&\textsf{3$\times$400}	&\textsf{1.04}\\
			&\textsf{B}		&\textsf{21/12/98}	&\textsf{3$\times$200}	&\textsf{1.12}\\
			&\textsf{V}		&\textsf{21/12/98}	&\textsf{3$\times$200}	&\textsf{0.92}\\
			&\textsf{R}		&\textsf{21/12/98}	&\textsf{3$\times$200}	&\textsf{0.84}\\\hline

			\textsf{NGC 4394}	&\textsf{U}		&\textsf{17/01/99}	&\textsf{2$\times$800,1$\times$400}	&\textsf{1.12}\\
			&\textsf{B}		&\textsf{18/01/99}	&\textsf{3$\times$200}	&\textsf{1.06}\\
			&\textsf{V}		&\textsf{18/01/99}	&\textsf{3$\times$200}	&\textsf{0.90}\\
			&\textsf{R}		&\textsf{18/01/99}	&\textsf{3$\times$200}	&\textsf{0.92}\\\hline\hline
\end{tabular}
\end{center}
\end{table}

As mentioned in Paper I, the flat fields
(one for each night and for each band) may be contaminated by scattered light
from the lack of baffling of the 3.5m APO telescope,
leading to an artificial spatial
gradient in the sky background. This is most evident in the B and R band
images than in the U and B ones. The one exception is NGC 4394 where some of
the gradient in the background is due to the presence of the galaxy M85, just
off the field of view.

Calibration was computed using observation of several standard stars
from Landolt (1992). An aperture magnitude for each observation of each
standard star was obtained with an
aperture radius of $7\arcsec$, chosen to match the aperture used by Landolt.
Standards
included 3 stars in the field of PG 1323-086 and 5 stars in the field of PG
1633+099, with color index in the range from -0.9 to 1.1 for U-B, from -0.2 to
1.1 for B-V and from -0.1 to 0.6 for V-R. The standards were mostly observed
at low airmass, less than 1.2, with the entire range being from 1.1 to 2.2.

\section{Results}
\subsection {NGC 3377A}

NGC 3377A is an almost face-on dwarf spiral galaxy, about $7 \arcmin$
north-west from the E6 giant elliptical NGC 3377, in the Leo I (M96)
group. The radial velocity difference of 120 km/s (de Vaucouleurs et al. 1991, hereafter RC3)
between these two galaxies suggests that NGC 3377A may be a
companion of the giant elliptical.
According to Tonry et al. (1997), NGC 3377A is about
10.7 Mpc distant, implying a linear separation from NGC 3377 of only
20 kpc. Sandage et al. (1991) and Knezek et al. (1999) showed that
this very low surface brightness galaxy has various peculiarities. In
particular, in spite of its high M$_{HII}$/L$_B$ ratio ($\sim 0.30$ in
solar units, which is a typical value for a Sc type spiral), it has a
very low star formation rate, about 0.003 M$_{\sun}$/yr, so that star
formation can continue for more than a Hubble time (Knezek et al. 1999).

The application of our algorithm to this galaxy led us to the
identification of 83 YSGs, whose positions in the galaxy are shown in
Fig. 1 and whose main characteristic parameters are given in Table 2.
{\bf
We remind that, in order to reduce the contamination by random groups in this list of YSGs candidates,
we adopted (as in Paper I) the procedure described by Battinelli \& Demers (1992). This procedure
results into the introduction of a threshold, N$_{lim}$, in the number of pixels in each groups, such
that the contamination by random groups in the sample of all the groups
composed by at least N$_{lim}$ pixels is
less than 10$\%$. The computed N$_{lim}$ are 9, 4, 7 for NGC 3377A, NGC 3507 and NGC 4394, respectively. 
}
Blue magnitudes and surface brightnesses have been corrected for a total
extinction of 0.06 mag (0.04 mag galactic extinction and 0.02 mag internal
extinction) as given in RC3.
As explained in Paper I (Sect. 3), a precise determination of integrated colours
of YSGs results hard since they often lie in areas with a lot of "structure".

The size distribution of the YSGs is shown in Fig. 4 for the three
galaxies studied in this paper. For NGC 3377A the average size is 87 pc
with standard deviation 59 pc. { \bf Fitting the high-size 
tail of the size distribution with the power law $dN\propto D_{xy}^\alpha dD_{xy}$
we find $\alpha=-2.3\pm0.3$ (correlation coefficient $r=0.93$.
The shaded strips shown in Fig. 4 are the D$_{xy}$ intervals directly effected
by the introduction of the N$_{lim}$ threshold for each galaxy. Such shaded areas
are therefore certainly uncomplete.}

Knezek et al. (1999) give a multi-color map of NGC 3377A where several
H$\alpha$ "knots" are evident over the whole optical disk.
A comparison of our YSG map with the H$\alpha$ knots shows
that a large percentage of the latter (about $80\%$) correspond to YSGs. Two out of the six H$\alpha$
sources that have no counterpart in our YSG sample are very close to a bright field star, three
are in the very outskirts of the optical galaxy and just one lies along a spiral branch.
The latter is a very tiny H$\alpha$ region perhaps related to a YSG too small to be resolved by us.
The most significant difference between our YSGs and
Knezek 's sources is in two big clumps near to the center that don't show well localized
H$\alpha$ emission, but are well characterized in our work.

\begin{figure}[ht]
\label{f1}
\resizebox{\hsize}{!}{\includegraphics[angle=0]{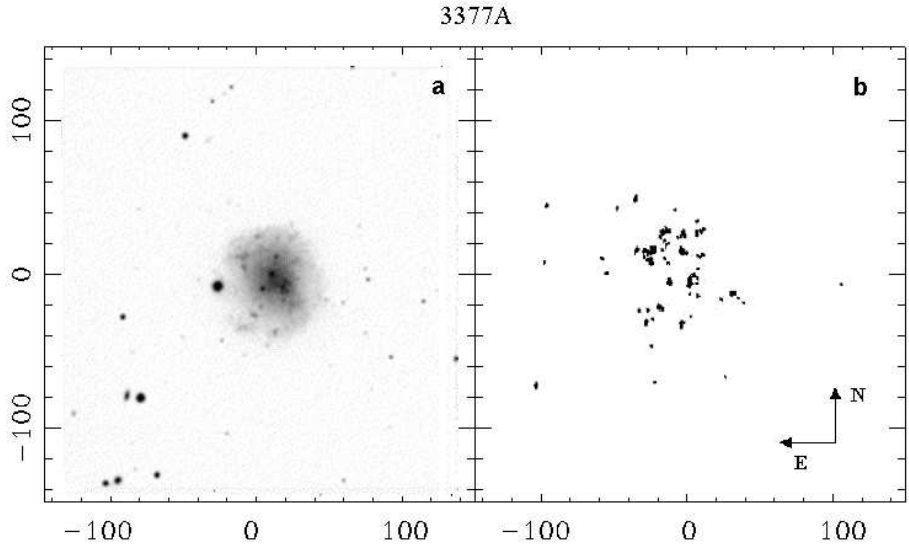}}
\caption{Panel {\bf a}) B band image of NGC 3377A. Panel {\bf b}) Map of the identified
YSGs. North is up and East to the left. Coordinates are in arcseconds and the
offset is relative to the galactic centre}
\end{figure}

\begin{figure}[ht]
\label{f2}
\resizebox{\hsize}{!}{\includegraphics[angle=0]{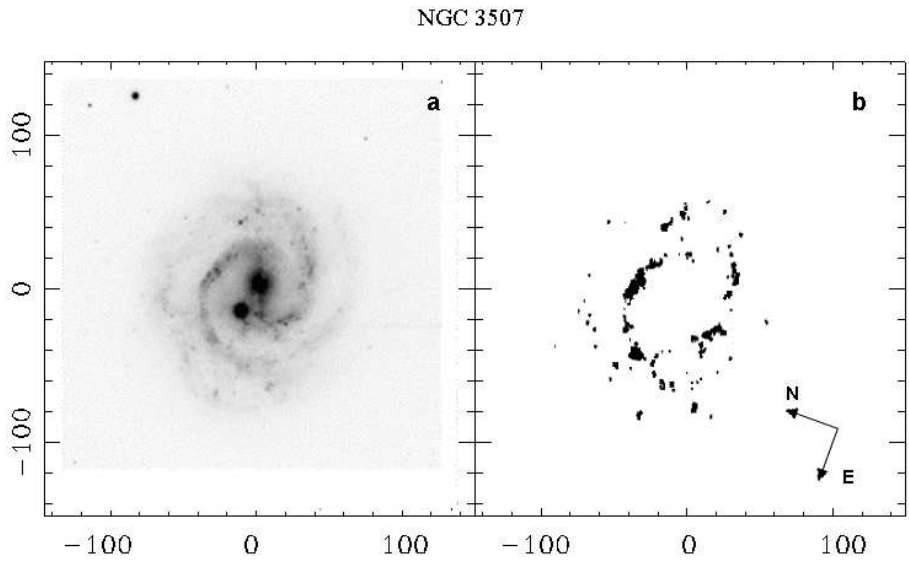}}
\caption{Panel {\bf a}) B band image of NGC 3507. Panel {\bf b}) Map of the identified
YSGs. Orientation of the image is shown. The angle between y-axis and
North is 70$^\circ$. Coordinates are in arcseconds and the offset is relative to the
galactic centre}
\end{figure}

\subsection {NGC 3507}
The spiral NGC 3507 is a typical barred galaxy (SBb type), with two long regular symmetric
arms circling around the nucleus and the bar. The inner parts of each arm are more regular
than the outer parts. The isophotal major axis is $D_{25}=3.4 \arcmin$ (RC3), the inclination is
26 $^\circ$ (Schulman et al. 1997) while the position angle is about 90 $^\circ$
(de Jong \& van der Kruit 1994).
According to Garcia et al. (1993), NGC 3507 is the brightest member of a small group
of galaxies. From the radial velocity of 906 km/s (RC3) and $H_0$=75 km/s/Mpc, a distance
of 12.1 Mpc is obtained, that corresponds to a scale of 16.4 pc/pixel.
Schulman et al. (1997) studied the HI distribution and dynamics in a small sample of galaxies, including NGC 3507.
This paper concludes that in NGC 3507 there are no High Velocity Clouds (HVCs,
i.e. HI regions with peculiar motion). Since HVCs are thought to be star formation indicators
(being probably originated by Supernova explosions) the lack of
HVCs in NGC 3507 suggests a low star formation rate in this galaxy.
Fig. 2 shows the map of the 90 YSGs detected: most of the objects are located along the inner part
of the arms. In Table 3 we give the main properties of the YSGs.
Galactic and internal B-absorptions are taken 0.00 mag and 0.10 mag, respectively, from RC3.
The YSGs average size is 121 pc, with a standard deviation of 147 pc
{ \bf and the slope of the fitted power law to the size distribution is
the smallest one with $\alpha=-1.6\pm0.3$ ($r=0.80$).}

\subsection {NGC 4394}
NGC 4394 is a barred and ringed spiral galaxy, with integrated B
magnitude 11.53 mag and with $D_{25}=3.6 \arcmin$ (RC3). Its
proximity to M85, just $6\arcmin$ East, and their similar radial
velocities suggest that NGC 4394 may belong to the M85 subgroup of the
Virgo Cluster. Thus we assume as distance of this galaxy the same (about 16 Mpc)
deduced by Ferrarese et al. (2000) for the M85 subcluster basing on Cepheid's indicators.
The inclination and the position angle are taken from Chapelon et al. (1999) and are
25$^\circ$ and 103$^\circ$, respectively.

The distribution of the YSGs, shown in Fig. 3, is clearly concentrated
in the ring region and just few YSGs are detected along the spiral arms.
This is supported by the radial distribution of
the surface luminosity density, $\sigma_B$, defined as the total B
luminosity of the YSGs found in an annulus divided by its area (see Fig. 5).
{\bf This quantity is clearly related to the high--mass star formation rate.
The concentrated distribution of the YSGs} probably reflects the high
abundance of cold gas detected in the rings of galaxies (Wong et
al. 2000). Moreover, HII regions were studied by Hodge (1974) in NGC
4394 with a 2.1 m telescope. He also found a clear concentration
of the HII regions in the ring, few of them being distributed in the
outer arms.

Table 4 gives the YSG parameters; the average
YSG diameter results to be 114 pc with a standard deviation of 77 pc,
{\bf while the exponent $\alpha$ of the power law is $-2.7\pm0.3$ ($r=0.92$).}

\begin{figure}[ht]
\label{f3}
\resizebox{\hsize}{!}{\includegraphics[angle=0]{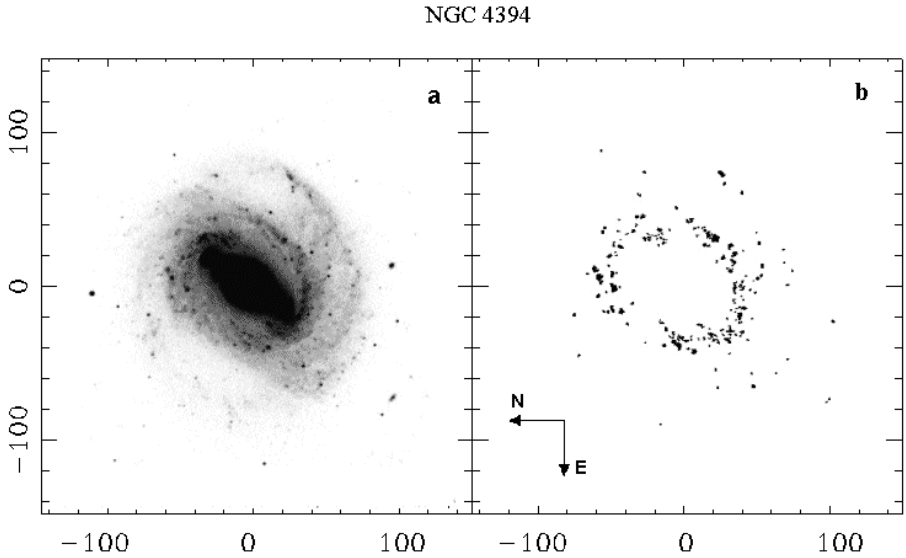}}
\caption{Panel {\bf a}) B band image of NGC 4394. Panel {\bf b}) Map of the identified
YSGs. North is left and East is down. Coordinates are in arcseconds and
the offset is relative to the galactic centre}
\end{figure}

\begin{figure}[ht]
\label{f4}
\resizebox{\hsize}{!}{\includegraphics[angle=0]{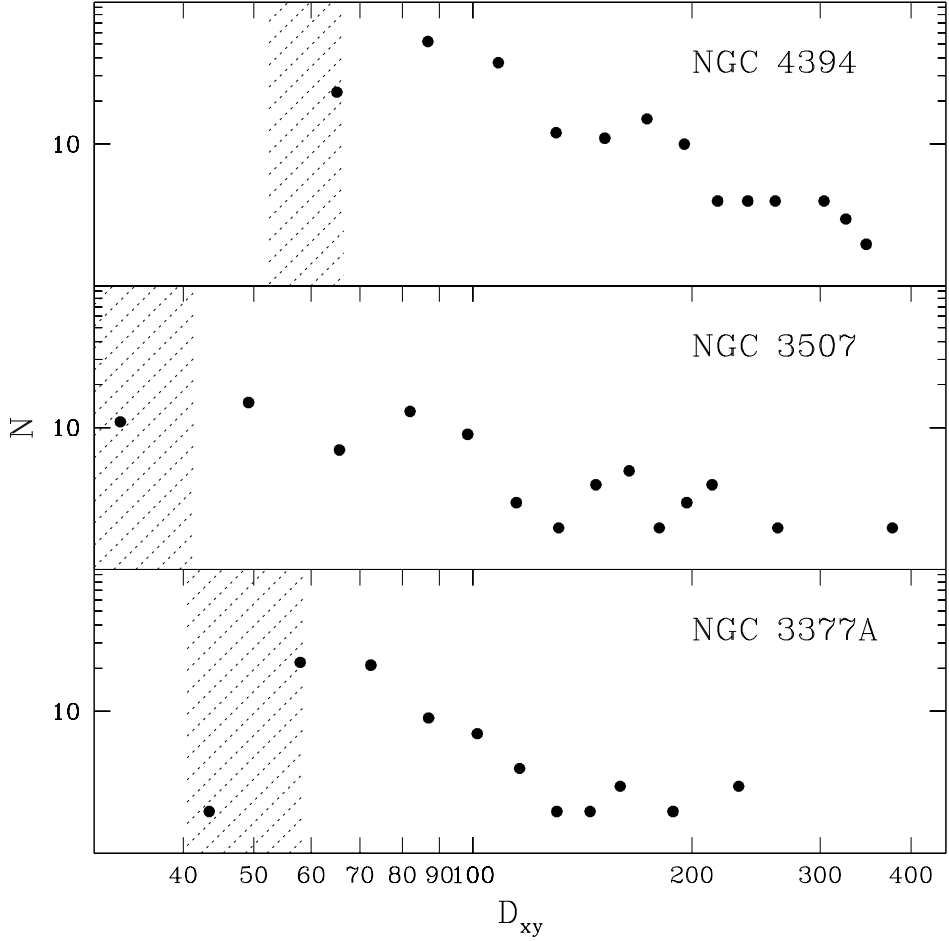}}
\caption{Size distribution of the YSGs. The size (in pc) is defined as the average
of the x and y extents of the YSG. Shaded areas are the uncompleteness strips as described in the
text.}
\end{figure}

\begin{figure}[ht]
\label{f5}
\resizebox{\hsize}{!}{\includegraphics[angle=0]{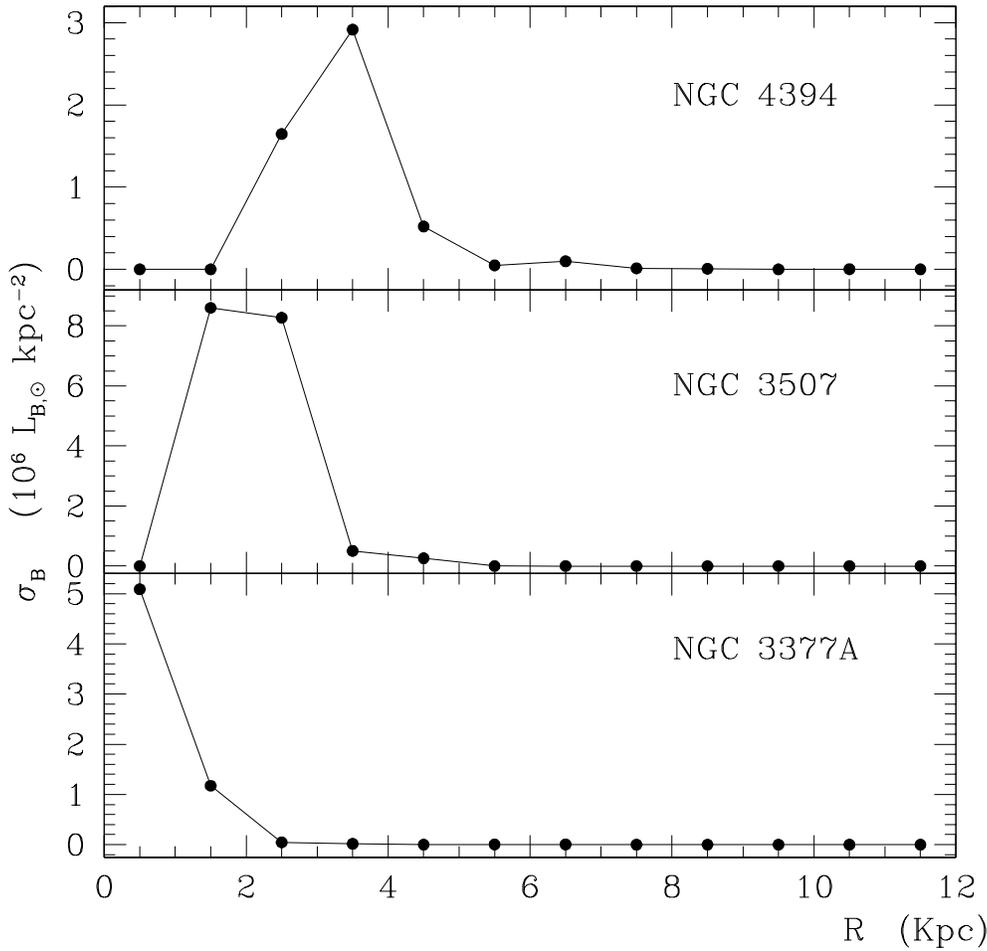}}
\caption{The surface luminosity density
($\sigma_B$, in $10^6 L_{B,\odot}$ kpc$^{-2}$) of YSGs averaged over
concentric circular annuli as a function of the galactrocentric radius (R, in kpc)
in NGC 3377A, NGC 3507 and NGC 4394. The YSG positions were corrected
for the galaxy inclination.}
\end{figure}

\begin{figure}[ht]
\label{f6}
\resizebox{\hsize}{!}{\includegraphics[angle=0]{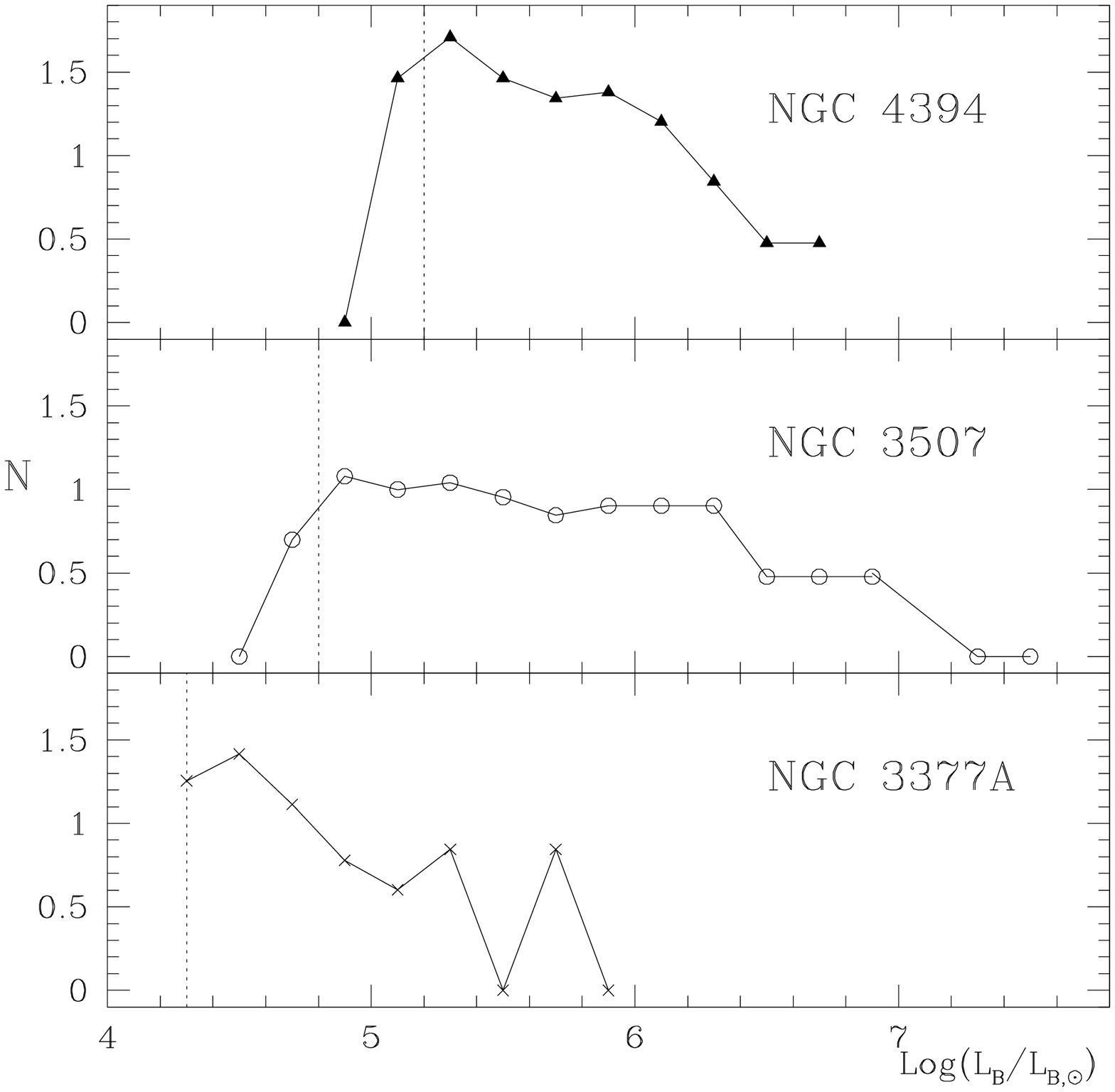}}
\caption{The differential luminosity function of the YSGs in B-band. Crosses, circles and triangles
refer to NGC 3377A, NGC 3507 and NGC 4394, respectively.}
\end{figure}

\section {Discussion and conclusions}
{ \bf With our semi-automatic identification algorithm we have determined the population and main
characteristics of Young Star Groupings in the spiral galaxies NGC 3377A, NGC 3507 and NGC 4394.
The data collected constitute not only an enlargement of the YSG data-base but also a direct source of
astrophysical information about the properties of the regions of star formation in spirals.

The analysis of the YSG differential B luminosity functions (Fig. 6)
shows that their high luminosity tails can be well fitted with power laws $dN\propto L_B^\beta dL_B$.
We find $\beta=-1.79\pm0.10$, $-1.31\pm0.08$ and $-1.90\pm0.18$, for NGC 3377A, NGC 3507 and NGC 4394,
respectively (the correlation coefficients r are equal to 0.94, 0.99 and 0.99).
These values are in good agreement to those found by Elmegreen \& Salzer (1999) for the star forming
complexes in a sample of 11 galaxies.
Again, the introduction of the N$_{lim}$ threshold described in Sect. 3.1 implies a correspondent detection
limit in the YSG luminosity. We evaluated such detection limits for each galaxy, by plotting the
luminosity as a function of number of pixels of the YSGs and then determining the value of the luminosity
corresponding to N$_{lim}$.
The detection limits given in Fig. 6
are close to the peaks of the luminosity functions, implying statistical uncompleteness in those regions.}
%The peculiarly low peak-luminosity in NGC 3377A was expected due to the very low surface brightness
%of this dwarf spiral.

The sample of six spirals studied so far (three in Paper I and three in this paper) allows
us a preliminary investigation of the existence of correlations among the YSG and parent galaxy properties.
In Fig. 7 we show some evidence of positive correlations among the number, dimension,
integrated B magnitude and the total area of the YSG populations with global
characteristics of the galaxies (integrated B magnitude and dimension).

\begin{figure}[ht]
\label{f7}
\resizebox{\hsize}{!}{\includegraphics[angle=0]{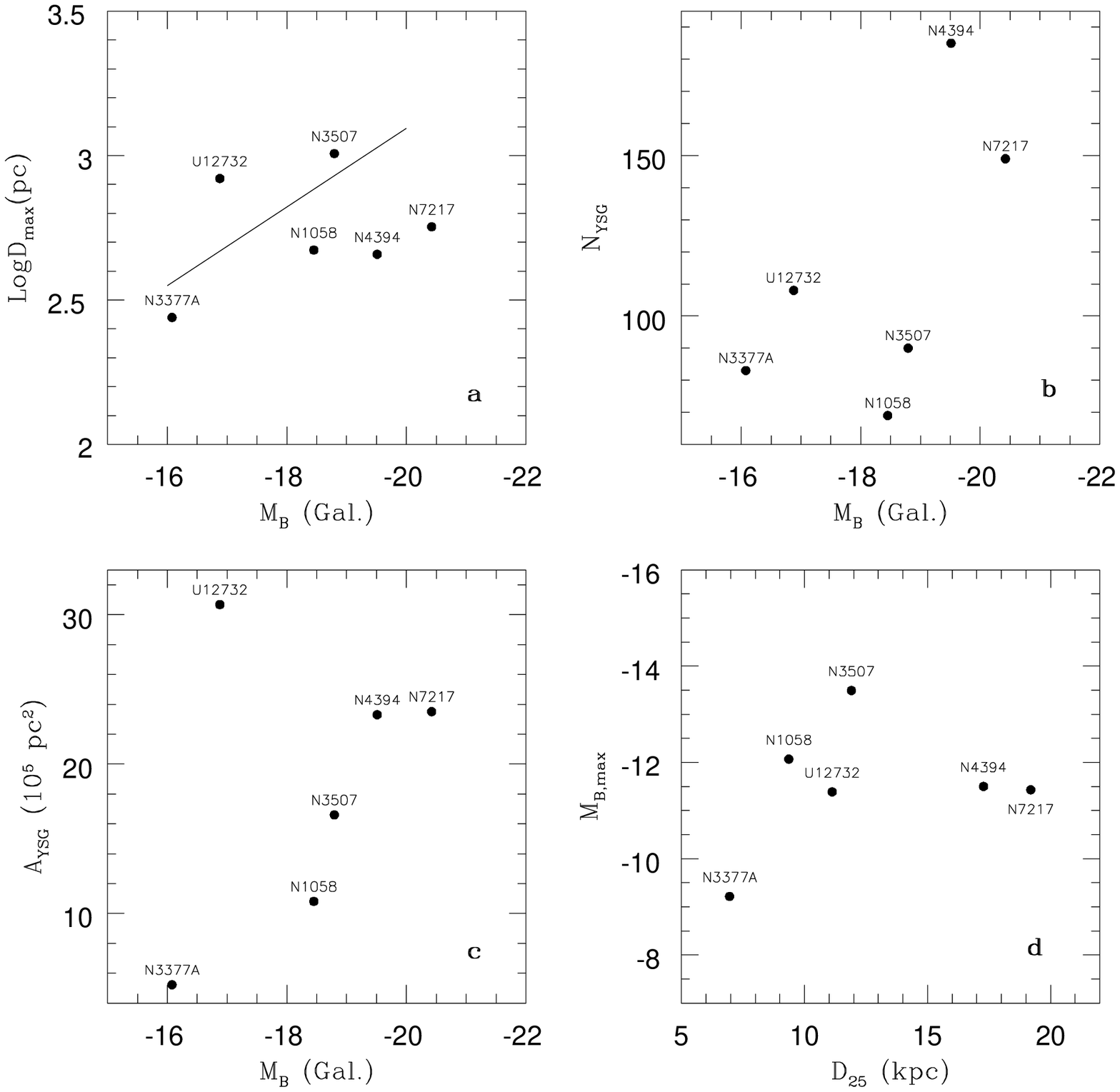}}
\caption{Correlations between YSG and parent galaxy properties. Panel {\bf a})
Logarithm of the
size of the largest YSG vs the integrated B magnitude of the galaxy (M$_B$); the straight line is the
Elmegreen et al. (1994) relation (see text). Panel {\bf b}) Number of YSGs
($N_{YSG}$) vs M$_B$. Panel {\bf c}) Total area of the YSG ($A_{YSG}$)
vs $M_B$. Panel {\bf d}) Blue magnitude of the brightest YSG
(M$_{B,max}$) vs the diameter of the galaxy (D$_{25}$) taken from RC3.}
\end{figure}

Incidentally, we note that in Fig. 7 the two ringed galaxies, NGC 7217 and
NGC 4394, are close each other and, at least in panel {\bf b}) and {\bf d}),
they seem to behave differently from the rest of the sample. Of course,
a larger number of ringed galaxies is necessary to check how real is such a
behaviour. In particular, the presence of these two galaxies
in our small sample results into a shallower slope of
the correlation between the size of the largest YSG
in a galaxy and the galaxy $M_B$ respect to the one obtained by
Elmegreen et al. (1994), who claimed their correlation closely matching
the expected variation in the characteristic length of the gaseous gravitational instability
with $M_B$.
{\bf Such correlation  has been suggested by Selman \& Melnick (2000) as a mere result of size-of-sample
effect acting on a universal size distribution ($dN \propto D_{xy}^{-4.2} dD_{xy}$). In order to check
the Selman \& Melnick (2000) suggestion  we can compare the  $-4.2$ slope with those of the fitted power laws
to the size distributions of the YSGs in NGC 3377A, NGC 3507 and NGC 4394, limited to the statistically
reliable size ranges. The slopes found ($-2.3$,$-1.6$,$-2.7$ for NGC 3377A, NGC 3507, and NGC 4394,
respectively) are quite close to the value, $-2$, for star forming molecolar cloud size distribution
(see Solomon et al. 1987 and Larson 1981). They are always too shallow to explain the Elmegreen et al. (1994)
relation by mean of the size-of-sample effect. We stress, however, that our six points in Fig. 7a
suggest a slope not as steep as Elmegreen et al's relation.
Anyway, we remark that our sample is yet too small to draw firm conclusion about this question.}

%\begin{acknowledgements}
%Thanks
%\end{acknowledgements}

%\documentclass[article]{aa}
%\begin{document}

\begin{table*}[ht]
\begin{center}
\label{tab1}
\setlength{\tabcolsep}{1.2mm}
\caption{YSGs of NGC 3377A. N is the identification number, X and Y are the
coordinates in arcsecs. R is the galactocentric distance (in kpc), corrected for the inclination
of the galaxy. B is the total blue magnitudes. $\Sigma _{B}$ is the surface brightness
(in mag/arcsec$^2$). Magnitudes are corrected by the extinction, from RC3.
D$_{xy}$ is the linear dimension (in pc).}
\begin{tabular}{|ccccccc||ccccccc|}
\hline
N & X & Y & R & B & $\Sigma_{B}$ & D$_{xy}$ & N & X & Y & R & B & $\Sigma_{B}$ & D$_{xy}$ \\
\hline
1&39&-127&6.9&24.56&24.29&72&43&-26&12&1.51&23.65&24.14&86\\
2&99&-121&8.13&24.67&24.3&43&44&11&11&0.84&23.66&24.09&86\\
3&-130&-96&8.42&24.4&24.33&72&45&-163&12&8.49&23.22&23.15&57\\
4&-136&-83&8.27&24.43&24.37&72&46&9&12&0.83&24.19&24.21&72\\
5&-166&-82&9.61&24.56&24.5&72&47&12&13&0.97&23.44&24.37&115\\
6&-99&-71&6.35&22.07&23.55&159&48&-21&15&1.37&21.08&23.59&275\\
7&-19&-68&3.71&24.64&24.58&57&49&-32&14&1.86&24.36&24.1&72\\
8&63&-66&4.76&24.77&24.39&57&50&-1&15&0.83&21.13&23.63&231\\
9&28&-65&3.69&24.65&24.39&57&51&2&14&0.78&23.74&24.52&101\\
10&-122&-49&6.84&24.78&24.52&72&52&-2&14&0.78&24.91&24.53&57\\
11&-21&-45&2.62&23.21&23.64&86&53&-26&15&1.59&21.84&23.56&173\\
12&-1&-31&1.63&21.43&23.63&231&54&-31&16&1.85&22.57&24.03&144\\
13&-25&-31&2.1&24.58&24.6&72&55&-14&16&1.15&23.97&24.29&86\\
14&-24&-29&2&23.46&23.84&86&56&-25&17&1.58&24.47&24.31&72\\
15&-21&-28&1.84&24.48&24.1&57&57&-21&17&1.46&22.71&23.93&130\\
16&4&-26&1.39&24.28&24.22&72&58&-19&18&1.4&24.1&24.03&72\\
17&-29&-22&1.93&24.27&24.29&72&59&-12&18&1.15&24.52&24.14&57\\
18&-24&-22&1.72&24.04&24.21&72&60&-10&20&1.21&24.1&24.48&101\\
19&-15&-20&1.35&21.38&23.83&231&61&-12&22&1.34&23.57&24.21&101\\
20&40&-17&2.28&25.02&24.64&57&62&-3&24&1.28&23.73&24.46&115\\
21&25&-15&1.53&23.71&24.2&86&63&-15&24&1.52&22.86&24.25&130\\
22&36&-14&2.03&24.02&23.95&57&64&0&24&1.29&24.76&24.49&57\\
23&9&-13&0.84&23.08&23.72&101&65&-1&25&1.34&24.13&24.31&72\\
24&33&-11&1.82&22.31&23.93&159&66&0&27&1.42&22.23&24.1&188\\
25&4&-11&0.65&24.18&23.91&57&67&8&25&1.4&24.6&24.53&72\\
26&-161&-8&8.35&24.27&24.58&72&68&-12&28&1.61&21.29&23.8&260\\
27&3&-4&0.31&21.41&24.08&289&69&8&27&1.47&23.95&24.05&72\\
28&-9&-4&0.54&20.87&23.34&217&70&9&28&1.53&23.75&24.13&101\\
29&106&-5&5.5&23.86&23.79&57&71&12&29&1.64&22.5&24.07&188\\
30&8&-5&0.53&25.01&24.64&43&72&-11&29&1.65&24.51&24.13&57\\
31&9&-1&0.49&25&24.62&57&73&-12&31&1.73&24.62&24.24&57\\
32&5&0&0.3&24.23&24.33&86&74&9&35&1.87&23.67&24.4&101\\
33&7&0&0.39&24.44&24.37&72&75&-44&43&3.24&22.96&23.89&115\\
34&7&0&0.36&24.73&24.35&57&76&-5&42&2.22&24.23&24.26&72\\
35&-52&1&2.7&22.51&24.04&144&77&-92&45&5.33&22.27&23.52&115\\
36&9&4&0.56&24.89&24.62&72&78&-32&49&3.07&22.23&23.82&159\\
37&-22&8&1.25&21.1&23.61&246&79&-118&94&7.85&24.88&24.61&57\\
38&-10&8&0.69&24.54&24.48&72&80&-154&95&9.4&24.99&24.73&57\\
39&-94&8&4.89&24.22&24.15&57&81&-149&115&9.8&24.13&24.44&101\\
40&-54&10&2.9&23.32&24.1&86&82&35&125&6.73&24.37&24.11&57\\
41&-12&10&0.87&24.29&24.66&86&83&77&130&7.85&24.21&24.05&57\\
42&-21&11&1.27&24.79&24.53&57
 &  &  &  &  &  &  & \\\hline
\end{tabular}
\end{center}
\end{table*}

\begin{table*}
\begin{center}
\label{tab2}
\setlength{\tabcolsep}{1.2mm}
\caption{YSGs of NGC 3507. See caption of Table 1.}
\begin{tabular}{|ccccccc||ccccccc|}
\hline
N & X & Y & R & B & $\Sigma_{B}$ & D$_{xy}$ & N & X & Y & R & B & $\Sigma_{B}$ & D$_{xy}$ \\
\hline
1&-30&-75&4.76&19.88&22.26&229&46&-65&-17&4.26&21.75&22.3&98\\
2&17&-76&4.77&22.67&22.69&81&47&54&-17&3.73&22.18&22.67&81\\
3&5&-72&4.34&20.3&22.31&196&48&-63&-12&4.13&21.57&22.38&114\\
4&6&-69&4.19&20&22.19&212&49&-71&-10&4.6&22.8&22.53&65\\
5&-7&-59&3.56&21.88&22.37&98&50&30&-10&2.11&23.5&22.48&32\\
6&-14&-58&3.51&21.03&22.33&147&51&-40&-9&2.64&23.58&22.32&32\\
7&4&-58&3.52&22.7&22.53&65&52&-37&-5&2.44&23.4&22.38&32\\
8&-16&-58&3.54&22.77&22.61&81&53&-36&-5&2.38&23.52&22.5&48\\
9&-8&-56&3.35&20.53&22.23&163&54&-67&-4&4.34&24.02&22.77&32\\
10&-15&-54&3.35&20.81&22.34&163&55&-33&5&2.2&16.82&21.92&1032\\
11&8&-55&3.36&21.74&22.33&81&56&28&0&1.86&19.69&22.22&294\\
12&4&-55&3.31&23.8&22.54&32&57&-38&-2&2.45&23.49&22.24&32\\
13&-49&-53&4.36&21.84&22.48&98&58&-27&3&1.77&23&22.49&48\\
14&10&-52&3.26&22.54&22.47&65&59&34&9&2.25&18.94&22&376\\
15&-20&-49&3.14&21.29&22.38&163&60&28&10&1.92&21.95&22.12&81\\
16&1&-49&2.95&23.97&22.95&48&61&32&11&2.13&21.62&22.26&98\\
17&21&-49&3.31&23.18&22.67&48&62&-63&11&4.22&22.87&22.71&65\\
18&-36&-46&3.48&23.31&22.8&65&63&-27&13&1.97&22.56&22.18&65\\
19&-46&-44&3.84&20.28&22.4&212&64&33&14&2.24&20.01&22.11&196\\
20&-21&-43&2.83&20.56&22.33&180&65&-25&15&1.97&22.9&22.39&48\\
21&-32&-36&2.94&17.75&22.1&622&66&-20&18&1.77&18.38&22.09&491\\
22&-26&-40&2.86&21.39&22.31&98&67&32&18&2.29&20.39&22.16&180\\
23&30&-40&3.2&21.68&22.37&98&68&33&17&2.35&22.47&22.31&81\\
24&-19&-38&2.55&21.21&22.43&130&69&5&20&1.24&23.13&22.47&48\\
25&10&-39&2.48&22.46&22.56&81&70&1&22&1.33&21.81&22.18&81\\
26&11&-36&2.33&20.29&22.22&212&71&-13&24&1.73&21.28&22.17&98\\
27&-14&-36&2.3&23.81&22.79&48&72&3&24&1.46&20.19&22.11&163\\
28&-10&-36&2.2&23.51&22.7&32&73&31&24&2.41&20.38&21.97&147\\
29&13&-34&2.27&22.44&22.61&81&74&0&24&1.48&22.48&22.32&65\\
30&10&-31&2.05&20.44&22.18&163&75&-3&25&1.53&22.73&22.07&48\\
31&-59&-32&4.19&21.95&22.59&98&76&22&27&2.1&22.03&22.4&81\\
32&-86&-32&5.78&24.33&23.08&32&77&29&29&2.51&21.68&22.32&114\\
33&6&-29&1.83&23.01&22.19&48&78&-1&34&2.07&23.32&22.81&48\\
34&6&-27&1.72&19.53&22&262&79&37&36&3.14&20.58&22.17&147\\
35&-36&-28&2.81&23.2&22.38&48&80&-13&42&2.71&18.85&21.76&278\\
36&15&-26&1.91&18.52&21.68&393&81&-9&44&2.78&20.18&22.11&196\\
37&20&-22&1.95&18.7&21.95&376&82&-50&45&4.39&21.18&22.33&114\\
38&25&-24&2.26&20.02&22.17&212&83&-39&45&3.82&23.32&22.5&32\\
39&14&-24&1.78&23.11&22.09&48&84&2&47&2.8&22.34&22.36&81\\
40&-39&-24&2.82&23.34&22.32&32&85&-1&49&2.97&19.51&22.18&262\\
41&-39&-19&2.73&18.97&22.18&475&86&26&51&3.38&20.52&22.18&147\\
42&-60&-22&4.03&21.24&22.2&98&87&0&54&3.21&21.25&22.31&130\\
43&-36&-20&2.56&23.06&22.41&48&88&0&56&3.37&23.34&22.52&48\\
44&-35&-19&2.49&22.26&22.43&81&89&15&57&3.5&23.37&22.55&48\\
45&22&-19&1.93&22.01&22.25&81&90&16&58&3.56&24.14&22.88&32\\
\hline
\end{tabular}
\end{center}
\end{table*}

\begin{table*}
\begin{center}
\label{tab3}
\setlength{\tabcolsep}{1.0mm}
\caption{YSGs of NGC 4394. See caption of Table 1.}
\begin{tabular}{|ccccccc||ccccccc||ccccccc|}
\hline
N & X & Y & R & B & $\Sigma_{B}$ & D$_{xy}$ & N & X & Y & R & B & $\Sigma_{B}$ & D$_{xy}$ & N & X & Y & R & B & $\Sigma_{B}$ & D$_{xy}$\\
\hline
1&50&-87&7.91&21.97&22.51&108&63&-35&-18&3.42&23.21&22.84&65&125&-33&28&3.54&23.48&22.83&87\\
2&-28&-73&6.29&20.35&22.83&347&64&-35&-15&3.29&23.25&22.88&87&126&7&30&2.49&21.59&22.93&173\\
3&21&-73&5.98&21.59&22.93&173&65&-36&-14&3.37&21.17&22.84&239&127&-34&31&3.72&20.91&22.63&304\\
4&-30&-66&5.89&22.14&22.78&130&66&-25&-14&2.49&23.53&23.15&87&128&-20&29&2.86&22.48&22.32&87\\
5&-42&-60&6.02&21.57&22.89&173&67&-69&-14&6.06&23.3&23.23&87&129&-29&30&3.38&22.55&22.65&130\\
6&26&-59&5.04&23.6&23.22&87&68&-55&-11&4.82&21.47&22.93&173&130&-36&31&3.91&23.43&22.92&87\\
7&24&-58&4.92&22.92&23.09&108&69&45&-11&3.91&22.64&22.95&108&131&-42&32&4.34&22.99&22.93&108\\
8&0&-50&3.92&21.87&22.84&152&70&-35&-11&3.2&23.47&22.96&65&132&-41&32&4.28&22.37&23.06&173\\
9&22&-44&3.9&20.99&22.77&239&71&54&-9&4.63&20.62&22.75&325&133&3&32&2.56&22.53&22.84&108\\
10&26&-44&4.04&21.66&22.87&173&72&-32&-10&2.93&23.2&23.04&108&134&-20&33&3.06&22.03&22.35&130\\
11&-8&-42&3.41&21.5&22.96&217&73&-34&-8&3.03&21.09&22.62&195&135&-22&33&3.18&21.31&22.34&195\\
12&-6&-41&3.28&21.86&22.93&195&74&-74&-8&6.43&23.03&23.05&87&136&-37&33&4.07&22.9&23&87\\
13&38&-42&4.54&23.82&23.32&87&75&42&-6&3.64&21.12&22.74&260&137&-34&33&3.86&22.71&22.89&108\\
14&26&-39&3.74&23.65&22.99&65&76&40&-7&3.43&23.42&22.77&65&138&-29&33&3.57&23.08&22.81&108\\
15&27&-38&3.7&23.29&23.02&87&77&59&-6&5.03&23.31&23.15&87&139&2&33&2.66&23.15&22.5&65\\
16&-17&-37&3.34&23.09&22.83&87&78&-54&-5&4.69&21.73&22.27&130&140&-24&34&3.29&22.8&22.42&87\\
17&47&-36&4.87&22.24&23.01&130&79&51&-4&4.38&19.55&22.32&347&141&-1&35&2.75&21.59&22.75&217\\
18&-13&-37&3.16&23.31&23.05&87&80&-44&-5&3.83&23.07&23&108&142&-25&35&3.4&23.04&22.67&87\\
19&-16&-36&3.25&23.55&23.05&87&81&43&-4&3.7&23.09&22.83&108&143&5&35&2.79&23.52&22.87&65\\
20&-10&-36&3.04&22.21&22.85&108&82&-52&-3&4.45&21.94&23.1&173&144&-40&35&4.33&23.12&23.06&87\\
21&12&-35&2.93&21.13&22.57&195&83&-49&-1&4.23&23.32&23.06&87&145&-8&35&2.82&23.25&22.87&87\\
22&5&-35&2.79&23.47&23.1&87&84&-42&-1&3.63&22.71&22.88&108&146&-22&35&3.3&23.05&22.68&65\\
23&-17&-35&3.19&22.33&22.87&130&85&50&0&4.26&21.91&22.98&173&147&-17&36&3.11&23.35&22.84&87\\
24&29&-34&3.59&21.48&22.61&195&86&-37&0&3.2&23.13&22.75&108&148&-3&36&2.82&22.2&22.74&130\\
25&38&-34&4.19&22.93&23.03&108&87&-37&1&3.17&21.99&22.72&195&149&3&35&2.82&23.37&22.99&65\\
26&48&-34&4.84&21.56&22.92&173&88&41&2&3.57&19.5&22.44&412&150&4&35&2.82&23.49&22.84&65\\
27&-20&-32&3.12&20.68&22.81&304&89&-71&0&6.1&22.59&22.9&108&151&-34&36&3.96&23.53&23.03&87\\
28&-51&-33&5.18&23.96&23.31&65&90&-41&1&3.5&21.83&22.42&152&152&-20&36&3.27&23.24&22.86&108\\
29&10&-33&2.71&22.7&22.72&108&91&37&2&3.23&23.24&22.98&87&153&-31&36&3.87&21.6&22.79&195\\
30&-24&-29&3.19&19.42&22.6&477&92&-42&3&3.6&21.07&22.73&260&154&0&36&2.86&23.51&22.86&65\\
31&19&-32&2.98&23.02&22.76&108&93&-37&3&3.17&21.88&22.61&173&155&-17&36&3.18&23.29&22.91&87\\
32&33&-33&3.73&23.11&22.73&87&94&-47&4&4&22.91&23.01&108&156&-13&36&3.06&23.22&22.84&65\\
33&12&-32&2.7&23.15&22.5&65&95&-36&5&3.16&21.72&22.57&173&157&-45&37&4.76&22.93&22.95&108\\
34&-12&-31&2.74&22.27&23.04&173&96&46&5&4.02&23.66&23.15&87&158&-24&37&3.51&22.32&22.5&108\\
35&17&-31&2.83&22.97&22.47&87&97&-50&6&4.32&22.66&22.9&108&159&-8&37&2.99&22.92&22.95&108\\
36&14&-31&2.69&22.63&22.36&87&98&52&8&4.51&21.41&22.89&217&160&-23&37&3.48&22.59&22.76&108\\
37&18&-30&2.81&21.96&22.39&130&99&-41&8&3.58&23.48&22.83&65&161&-3&37&2.9&23.48&22.97&87\\
38&22&-30&2.95&21.67&22.31&152&100&-36&11&3.18&22.04&22.47&152&162&-6&37&2.97&22.68&22.93&130\\
39&11&-30&2.5&22.96&22.58&87&101&-42&11&3.71&23.19&22.81&65&163&0&37&2.94&23.71&23.06&87\\
40&13&-29&2.56&23.25&22.59&87&102&29&12&2.73&23.57&23.06&87&164&0&40&3.12&20&22.78&456\\
41&9&-29&2.39&23.45&22.8&65&103&-37&12&3.28&22.86&22.48&65&165&-33&38&4.06&22.12&23.05&152\\
42&18&-28&2.69&21.83&22.47&152&104&30&13&2.83&22.68&22.92&108&166&-26&38&3.71&22.26&22.85&152\\
43&21&-28&2.83&22.91&22.54&87&105&42&16&3.93&21.18&23.04&282&167&-5&38&3.03&23.03&23.05&108\\
44&7&-29&2.33&23.44&22.79&87&106&-42&16&3.79&20.79&22.7&304&168&21&39&3.61&23.19&23.03&87\\
45&37&-29&3.82&23.2&22.7&65&107&-38&16&3.47&23.14&22.49&87&169&8&41&3.33&21.18&22.69&173\\
46&41&-27&4.07&22.83&23.01&108&108&36&18&3.49&20.72&22.88&304&170&-20&41&3.61&23.67&23.16&87\\
47&-53&-26&5.06&21.9&22.93&152&109&-35&17&3.23&23.08&22.57&87&171&-7&43&3.39&22.78&22.89&108\\
48&8&-26&2.2&21.69&22.42&152&110&45&18&4.21&22.46&23.05&108&172&-1&43&3.38&23.61&22.96&87\\
49&-17&-25&2.53&21.3&22.76&195&111&48&18&4.43&22.38&22.87&108&173&0&43&3.4&22.86&22.8&87\\
50&-34&-25&3.62&20.78&22.7&239&112&-37&19&3.48&23.11&22.46&65&174&-10&45&3.61&20.89&22.8&260\\
51&12&-25&2.2&22.16&22.18&87&113&-40&21&3.77&20.25&22.58&325&175&-33&46&4.53&23.73&23.08&65\\
52&-20&-24&2.65&22.78&22.96&130&114&67&20&6.03&21.5&22.6&152&176&64&47&6.78&22.35&22.99&130\\
53&37&-24&3.64&21.13&22.73&217&115&-32&22&3.22&22.95&22.3&87&177&-71&50&7.12&22.89&23.14&87\\
54&41&-25&3.98&22.6&22.85&108&116&-48&23&4.46&23.57&23.19&87&178&-24&57&4.84&22.7&23.24&108\\
55&-33&-23&3.45&23.35&22.98&87&117&-32&23&3.24&22.96&22.31&87&179&-64&59&7.03&23.45&23.19&65\\
56&-69&-22&6.21&22.56&23&108&118&-38&24&3.69&22.89&22.73&87&180&-48&59&6.09&23.03&23.13&108\\
57&-42&-21&4.08&22.05&21.54&87&119&-101&25&8.79&21.44&22.3&152&181&-49&68&6.65&21.08&22.66&195\\
58&-37&-19&3.61&20.93&22.79&260&120&12&24&2.23&23.46&23.08&108&182&-26&69&5.75&22.81&23.12&108\\
59&40&-20&3.77&23.49&22.84&65&121&-40&25&3.92&21.96&22.89&173&183&-99&76&10.13&22.75&22.68&108\\
60&44&-18&3.99&21.24&22.56&195&122&10&26&2.27&21.9&22.83&173&184&-97&78&10.1&23.38&23.01&87\\
61&40&-18&3.65&23.03&22.52&87&123&32&26&3.56&22.43&23.02&130&185&11&93&7.42&23.55&23.17&65\\
62&-29&-17&2.91&21.01&22.71&239&124&-15&28&2.48&20.63&22.33&325&  &  &  &  &  &  & \\\hline
\end{tabular}
\end{center}
\end{table*}

%\end{document}


\begin{thebibliography}{}
\bibitem{} Adanti, S., Battinelli, P., Capuzzo--Dolcetta, R., \& Hodge, P.W. 1994, A\&AS, 108, 395
\bibitem{} Battinelli, P., Capuzzo--Dolcetta, R., Hodge, P.W., Vicari, A., \& Wyder, T.K. 2000, A\&A, 357, 437
\bibitem{} Battinelli, P., \& Demers, S. 1992 AJ 104, 1458
\bibitem{} Chapelon, S., Contini, T., \& Davoust, E. 1999, A\&A, 345, 81
\bibitem{} de Jong, R., \& van der Kruit, P.C. 1994, AAS, 101, 451
\bibitem{} de Vaucouleurs, G., de Vaucouleurs, A., Corwin, H.G.Jr., et al. 1991, {\it Third Reference Catalogue of Bright Galaxies}, (Berlin:Springer-Verlag)
\bibitem{} Elmegreen, D.M., Elmegreen, B.G., Lang, C., \& Stephens, C. 1994, ApJ, 425, 57
\bibitem{} Elmegreen, B.G, Elmegreen, D.M., Salzer, J., \& Mann, H. 1996, ApJ, 467, 579
\bibitem{} Elmegreen, D.M., \& Salzer, J.J. 1999, AJ, 117, 764
\bibitem{} Ferrarese, L., Ford, C.H., Huchra, J., Kennicutt Jr., R.C., Mould, J.R., et al. 2000, ApJS, 128, 431
\bibitem{} Garcia, A.M. 1993, A\&AS, 100, 47
\bibitem{} Hodge, P.W. 1986, in {\it Luminous Stars and Associations in Galaxies}, IAU Symp. n.116 (Dordrecht:Reidel)
\bibitem{} Hodge, P.W. 1974, ApJS, 27, 113
\bibitem{} Knezek, P.M., Sembach, K.R., \& Gallagher, J.S. 1999, A\&AS, 514, 119
\bibitem{} Landolt, A.U. 1992 AJ, 104, 340
\bibitem{} Larson, R.B. 1981, MNRAS, 257, 119
\bibitem{} Sandage, A., \& Hoffman, G.L. 1991, ApJ, 379, L45
\bibitem{} Schulman, E., Bregman, J.N., \& Roberts, M.S. 1994, ApJ, 423, 180
\bibitem{} Selman, F.J., \& Melnick, J. 2000, ApJ, 534, 703
\bibitem{} Solomon, P.M., Rivolo, A.R., Barret, J., \& Yahil, A. 1987, ApJ, 319, 730
\bibitem{} Tonry, J.L., Blakeslee, J.P., Ajhae, E.A., \& Dressler, A. 1997, ApJ, 475, 399
\bibitem{} Young, J.S., Allen, L., Jeffrey, D.P., Lesser, A., \& Rownd, B. 1996, AJ, 112, 1903
\bibitem{} Wong, T., \& Blitz, L. 2000, ApJ, 540, 771

%\bibitem{} Bertelli, G., Bressan, A., Chiosi, C., Fagotto, F., Nasi, E. 1994, A\&AS, 106, 275
%\bibitem{} Bresolin, F., Kennicutt, R.C., Ferrarese, L., Gibson, B.K., Graham, J.A., Marci, L.M., Phelps, R.L., Rawson, D.M., Sakai, S., Silbermann, N.A., Stetson, P.B., Turner, A.M. 1998, AJ, 116, 119
%\bibitem{} Buta, R., van Driel,W., Braine, J., Combes, F., Wakamatsu, K., Sofue, Y., Tomita, A. 1995, ApJ, 450, 593
%\bibitem{} Buta, R., Crocker, D. 1991 AJ, 102, 1715
%\bibitem{} Capuzzo--Dolcetta, R., Vicari, A., Battinelli, P., Arrabito, G. {\it astro-ph/0101041}, Proceeding of Modes of Star Formations and the Origin of Field Populations workshop, in press
%\bibitem{} Efremov, Y.N., Ivanov, G.R., Nikolov, N.S. 1987, Ap\&SS, 135, 119
%\bibitem{} Elmegreen, D.M., Elmegreen, B.G. 1982, MNRAS, 201, 1021
%\bibitem{} Elmegreen, D.M., Elmegreen, B.G. 1987, ApJ, 314, 3
%\bibitem{} Ferguson, A.M.N., Gallagher, J.S., Wyse, R.F.G. 1998a, AJ, 116, 673
%\bibitem{} Ferguson, A.M.N., Wyse, R.F.G., Gallagher, J.S., Hunter, D.A. 1998b, ApJ, 506, L19
%\bibitem{} van den Bergh, S. 1964, ApJS, 9, 65
%\bibitem{} Verdes--Montenegro, L., Bosma, A., Athanassoula, E. 1995, A\&A, 300, 65
%\bibitem{} Wyder, T.K., Hodge, P.W., Battinelli, P., Capuzzo--Dolcetta, R. 1998, BAAS, 193.6004

% Interessante Astronomy Reports, Volume 45, Issue 1, January 2001, pp.1-15 da prendere
\end{thebibliography}
\end{document}